\documentclass[aps, pra, superscriptaddress, twocolumn, amsfonts, amsmath, amssymb, floatfix]{revtex4-2}
\usepackage{graphicx}
\usepackage{sidecap}
\usepackage{dcolumn}
\usepackage{bm}
\usepackage{epsfig}
\usepackage{braket}
\usepackage{color}
\usepackage{ulem}
\usepackage{amsmath}
\usepackage[toc,page]{appendix}
\usepackage[colorlinks=true, letterpaper=true, pdfstartview=FitV, linkcolor=blue, citecolor=blue, urlcolor=blue]{hyperref}
\begin{document}
\title{Spin Faraday Waves in Periodically Modulated Spin-Orbit-Coupled Bose Gases}

\author{Hongguang Liang}
\affiliation{Department of Physics, School of Physics and Electronic Science, East China Normal University, Shanghai 200241, China}

\author{Meiling Wang}
\affiliation{Department of Physics, School of Physics and Electronic Science, East China Normal University, Shanghai 200241, China}

\author{Juan Wang}
\affiliation{Department of Physics, School of Physics and Electronic Science, East China Normal University, Shanghai 200241, China}

\author{Yan Li}
\email{yli@phy.ecnu.edu.cn}
\affiliation{Department of Physics, School of Physics and Electronic Science, East China Normal University, Shanghai 200241, China}
\affiliation{Chongqing Key Laboratory of Precision Optics, Chongqing Institute of East China Normal University, Chongqing 401120, China}

\date{\today}
\begin{abstract}
This paper investigates the formation of Spin Faraday waves in spin-orbit-coupled Bose-Einstein condensate under the stripe phase and explores the dispersion relation under three different phases. We discover that the SFW exhibit temporal and spatial patterns when the interaction is modulated periodically, and appear with resonant waves and higher order harmonics. SFW can be excited even when the modulation frequency resonates with the trap frequency. Furthermore, we study the dispersion relation of these Faraday modes through periodic modulation, which agrees well with our theoretical results under three quantum phases. Our work indicates novel physical phenomena originating from the introduction of spin-orbit coupling and provides a possible method for studying the dispersion of Bose gases. 
\end{abstract}

\maketitle

\section{Introduction}
In 1831, Faraday observed standing waves on the surface of a liquid in a container subjected to forced oscillations. The frequency of the waves was half of the driving frequency, marking the origin of Faraday waves (FW). \cite{RN1}. The emergence of FW brings significant interest in driven systems, pattern formation, and other areas, leading to extensive research on FW in fluid systems, nonlinear optics, biological media, and chemical reactions \cite{RN2}. Theoretical studies have shown that the physical mechanism behind FW involve modulation instability or parametric resonance \cite{RN3}. These are both intriguing topics in physics research. Modulating the system and studying its excitation modes can help to understand the breakdown of spatial and temporal symmetries within the system. In 1995, the experimental creation of Bose-Einstein condensate (BEC) \cite{RN4} shifted the focus on FW research from classical liquids to quantum fluids, including both bosonic and fermionic systems. Considering the high controllability, BEC systems become an ideal platform for exploring nonlinear excitations in quantum fluids. This has led to numerous theoretical and experimental studies \cite{RN5,RN6,RN7,RN8,RN9,RN10,RN11}. FW have been observed in single-component cold atomic systems by periodically modulating the scattering length or confinement frequency \cite{RN5,RN6}. Experimental efforts have subsequently led to the creation of two-component BEC in two different hyperfine states. Coherent coupling between the two levels can be induced by microwave or radiofrequency waves, resulting in Rabi-coupled two-component BEC \cite{RN12,RN13,RN14}. These two-component BEC systems possess richer physical implications and promising applications, thus attracting considerable theoretical exploration \cite{RN15}.

In 2011, I. Spielman group from the U.S. National Institute of Standards and Technology successfully demonstrated the experimental realization of spin-orbit-coupled (SOC) BEC \cite{RN16}. This coupling between the spin and orbital motion of atoms is significant in condensed matter physics, contributing to various physical phenomena such as Spin Hall effects \cite{RN17,RN18} and topological insulators \cite{RN19,RN20}. The ground states of SOC BEC primarily consist of the stripe phase, plane phase, and zero momentum phase \cite{RN21}. The stripe phase is often used to study supersolid states. It is characterized by components with opposite momentum and the spontaneous breaking of translational symmetry and U(1) symmetry \cite{RN22,RN23}. Considerable research has focused on spontaneous pattern formation and parametric instability in SOC BEC, including the generation of FW through quenching process \cite{RN24,RN25,RN26,RN27,RN28}. However, studies on the excitation dynamics of Faraday patterns in SOC systems using periodic modulation are limited.

Inspired by experimental and theoretical works \cite{RN11,RN15,RN29}, in this paper we investigate the formation of Spin Faraday waves (SFW) in a spin-1/2 system, confined in a harmonic trap, specifically within the stripe phase. By periodically modulating the atomic scattering lengths through Feshbach resonance \cite{RN5,RN30,RN31,RN32}, SFW are excited at a typical frequency which is twice the stronger trap frequency. Different from non-SOC systems, SFW can also be generated when the driving frequency equals to trap frequency. We find that SFW remain stable only for a certain period of time and are accompanied with resonant waves and higher-order harmonics. Furthermore, we examine the dispersion relation of SFW in the stripe phase, plane wave phase and zero momentum phase. Our simulation results agree well with the prediction of the Bogoliubov-de Gennes (BdG) equations.

\begin{figure}[htbp]
	\centering
	\includegraphics[width=0.5\linewidth]{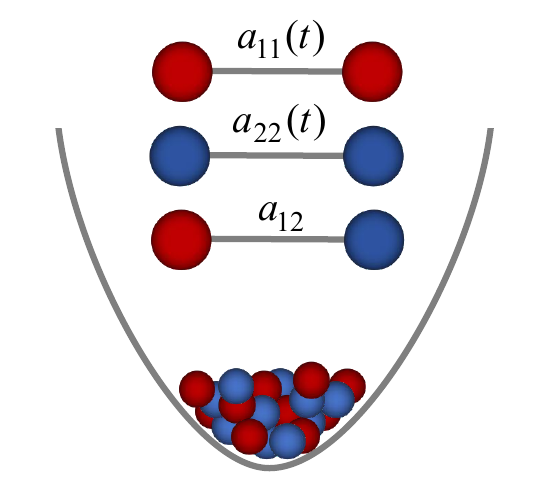}
	\caption{A schematic of atomic cloud localized in a 2D harmonic trap, where $a_{11}$ and $a_{22}$ are time-dependent, and $a_{12}$ is constant. Different colors represent atoms occupying different finestates.}
	\label{fig1}
\end{figure}

This paper is organized as follows: in the next section, we give the theoretical model and illustrate the modulation methods. In Sec. \ref{3} the SFW excited at two special modulation frequencies and the time evolution of SFW at a normal frequency are studied. In Sec. \ref{4} we study the dispersion relation of Faraday modes in different phases and compare them with BdG equations under plane wave phase and zero momentum phase. Finally, we give a conclusion of our work in the last section.

\section{Theoretical Model}
\label{2}

The model presented in this paper is based on the interaction between light and a three-level atomic system. We consider a two-dimensional, elongated $^{87}$Rb BEC confined in a harmonic trap  $V_{\text{trap}}=\frac{1}{2}m\omega^{2}_{x}x^{2}+\frac{1}{2}m\omega^{2}_{y}y^{2}$, where m is atomic mass, and $\omega_x$($\omega_y$) is trap frequency. A magnetic field is applied resulting in the splitting of Zeeman levels. Finally, two fine states $\ket{F=1,m_F=+1}$ and $\ket{F=1,m_F=0}$ are coupled by two counter-propagating Raman lasers interacting with the atomic cloud, while  $\ket{F=1,m_F=-1}$ is ignored for significant energy level difference \cite{RN16}. The two pseudospin states, $\ket{F=1,m_F=+1}$ and $\ket{F=1,m_F=0}$, represent the two components of spin up and spin down, respectively. One-dimensional spin-orbit coupling with equivalent contributions for Rashba and Dresselhaus is induced due to the Raman lasers. To constrain the degree of freedom in the z-direction, a sufficiently strong harmonic potential is applied, limiting the dynamics in the x and y directions. The single-particle Hamiltonian of the SOC BEC under the rotating wave approximation is as follows, as described in \cite{RN21,RN33,RN34}:
%\begin{equation}
\begin{flalign}
	\begin{split}
  H_{\text{sp}}=& \dfrac{\hbar^{2}}{2m}[(k_{x}-k_{0}\sigma_{z})^2+p_{\bot}^{2}]
         +\dfrac{\hbar\Omega}{2}\sigma_{x}+\dfrac{\hbar\delta}{2}\sigma_{z}+V_{\text{trap}}.
    \end{split}&
\end{flalign}
%\end{equation}
Here, $\hbar$ is the reduced Planck constant, $k_x$ and $p_{\bot}$ are momentum-space quantities, $k_0$ is the wave vector of the laser in x direction, $\Omega$ represents the Raman coupling strength, signifying the transition between the two atomic levels, $\delta$ is related to the energy difference between the two spin states, and $\sigma_{x,z}$ denote the corresponding Pauli matrix. Neglecting external potential and detuning, i.e., $V_{\text{trap}}=0$ and $\delta=0$. Diagonalizing $H_{\text{sp}}$ yields the eigenenergy of this single-particle system:
\begin{normalsize}
\begin{flalign}
&\	E_{k_{x},\pm}=\frac{\hbar^{2}k_{x}^{2}+p_{\bot}^{2}}{2m}+\frac{\hbar^{2}k_{0}^{2}}{2m}\pm\sqrt{\dfrac{\hbar^{4}k_{0}^{2}k_{x}^{2}}{m^2}+\frac{\hbar^{2}\Omega^{2}}{4}}.&
\end{flalign}
\end{normalsize}
As $\Omega$ increases, the eigenenergy $E_{k_{x},-}$ exhibits distinct structure.
	
	In the mean-field approximation, we use $\Psi_{1}$ and $\Psi_{2}$ to describe the wave functions of the two components, and define $k_0=\dfrac{\sqrt{2}\pi}{\lambda}$ and $E_r=\dfrac{\hbar k^{2}_0}{2m}$ as dimensionless units for length and frequency. The interaction Hamiltonian of this two-component system is
%	\begin{equation}
	\begin{flalign}
&\		H_{\text{int}}={\left(\begin{array}{cc}
				g_{11}|\Psi_{1}|^2+g_{22}|\Psi_{2}|^2 \qquad 0 \\[2mm]
				0 \qquad g_{22}|\Psi_{2}|^2+g_{21}|\Psi_{1}|^2
			\end{array}\right)}.&
		\end{flalign}
%	\end{equation}
	Here, $g_{ii}$ denotes the intraspecies interaction and $g_{ij}$ the interspecies interaction, $g_{ij}=4\pi\hbar^{2}a_{ij}/m$. And $a_{ij}$ are s-wave scattering lengths with $i, j$ the indices ($i,j={1,2}$) characterizing the two components. In the following section of this article, 1 represents spin up component and 2 spin down. The Hamiltonian of the many body system is $H=H_{\text{sp}}+H_{\text{int}}$. We can obtain the coupled GP equations for the SOC system \cite{RN24,RN35}:
		\begin{flalign}
%		i\hbar\frac{\partial}{\partial t}\Psi_{1}=-\frac{\hbar^{2}\nabla^{2}}{2m} \Psi_{1}+V_{pot}\Psi_{1}+(g'_{11} |\Psi_{1}|^{2}+g'_{12}|\Psi_{2}|^{2})\Psi_{1}+\frac{\Omega}{2}\Psi_{2} 
&\		 i\hbar\frac{\partial}{\partial t}\Psi_{1}=-\frac{\hbar^{2}(k_x-k_0)^2}{2m} \Psi_{1}+V_{\text{trap}}\Psi_{1}+A_1\Psi_{1}+\frac{\hbar\Omega}{2}\Psi_{2},& \label{gp1}   \\[2mm]
&\		 i\hbar\frac{\partial}{\partial t}\Psi_{2}=-\frac{\hbar^{2}(k_x+k_0)^2}{2m} \Psi_{2}+V_{\text{trap}}\Psi_{2}+A_2\Psi_{2}+\frac{\hbar\Omega}{2}\Psi_{1},& \label{gp2}
		\end{flalign}
	in which $A_1=g'_{11} |\Psi_{1}|^{2}+g'_{12}|\Psi_{2}|^{2}$, $A_2=g'_{22} |\Psi_{2}|^{2}+g'_{21}|\Psi_{1}|^{2}$ with $g^{'}_{ij}={g_{ij}}/(\sqrt{2\pi} a_z)$ the interaction coefficient in 2D system. The oscillator length is given by $a_z^2=\hbar/m\omega_z$.
%	\subsection{Modulation Method}

	We study the dynamical property of the system by periodically modulating the intraspecies interactions using Feshbach resonance. As depicted in Fig. \ref{fig1}, a bulk of $^{87}$Rb atom is confined in a harmonic trap, where the red circles represent atoms in spin up state and blue circles spin down state. Here, $a_{ii}$ denote the intraspecies scattering lengths, while $a_{ij}$ denote the interspecies scattering lengths. In this paper, we mainly adopt two modulation methods:  in-phase and out-of-phase modulation, as shown in the following equations:
	\begin{flalign}
% &\ a_{11}(t)=a_{11}+\Delta_{1} sin(\omega_{1}t),&  \label{in1}\\[2mm]
% &\ a_{22}(t)=a_{22}+\Delta_{2} sin(\omega_{1}t).& \label{in2}
 &\ a_{ii}(t)=a_{ii}+\Delta a \, \sin(\omega_{i}t).&  \label{in}
	\end{flalign}
	\begin{flalign}
%&\		a_{11}(t)=a_{11}+\Delta_{1} sin(\omega_{1}t),& \label{out1}\\[2mm]
%&\		a_{22}(t)=a_{22}+\Delta_{2} sin(\omega_{1}t+\pi),& \label{out2}
 &\ a_{ii}(t)=a_{ii}+\Delta a \, \sin(\omega_{i}t+(i-1)\pi).&  \label{out}
	\end{flalign}
	Eq. (\ref{in}) and Eq. (\ref{out}) indicate in-phase and out-of-phase modulation, respectively. $\Delta a$ is modulation amplitude, and $\omega_{i}$ represent modulation frequency of $\omega_{1}$ and $\omega_{2}$.
	
	Considering the time variation of the interaction $g_{ii}$, we numerically solve the coupled GP equations by evolving initial state with the time-evolution operator.
%	\begin{flalign}
% &\		\ket{\Psi(r,t)}=e^{-i\hat{H}t}\ket{\Psi(r,0)}.& \label{time}
%	\end{flalign}

	\begin{figure}[htb]
		\centering
		\includegraphics[width=1\linewidth]{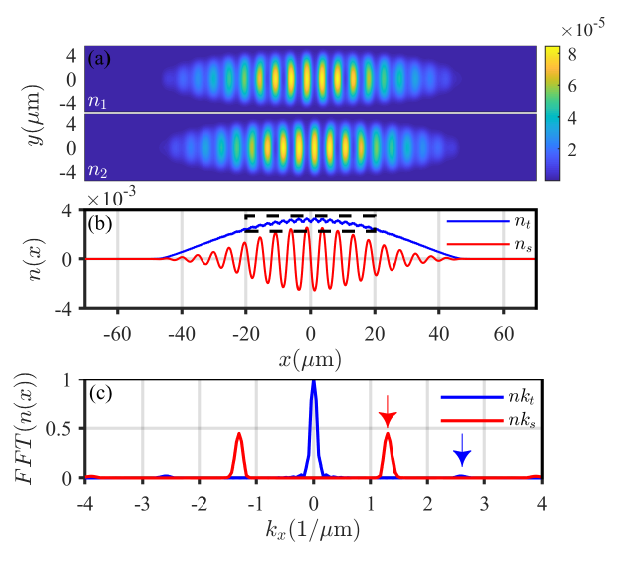}
		\caption{Spin Faraday patterns excited by in-phase modulation $\omega_i=200$ Hz and $\Delta a=6a_0$ in the stripe phase $\Omega=0.1E_r$.
			(a) Density distribution of two components in x-y direction.
			(b) Total density oscillation (black dash) and spin density oscillation (red line) that stands for the formation of SFW.
			(c) 1D Fourier transform of panel (b). The peaks pointed by the blue and red arrows represent density and spin Faraday waves, respectively.}
		\label{200}
	\end{figure}

	\section{Dynamic Analysis}
	\label{3}

	 $^{87}$Rb atoms are trapped in the potential with frequencies $\{ \omega_x,\omega_y,\omega_z\}=\{10,100,1000\}$ Hz, forming an elongated BEC along the x-direction. Our work numerically solves GP equations at different initial phase $\ket{\Psi(r,0)}$ and obtain $\ket{\Psi(r,t)}$. Considering the stripe phase, the wave function of ground state likes $1/\sqrt{2}\ket{\uparrow}+1/\sqrt{2}\ket{\downarrow}$. We excite the SFW directly using in-phase modulation, as shown in Fig. \ref{200}.

     \begin{figure}[htbp] 
		\centering
		\includegraphics[width=1\linewidth]{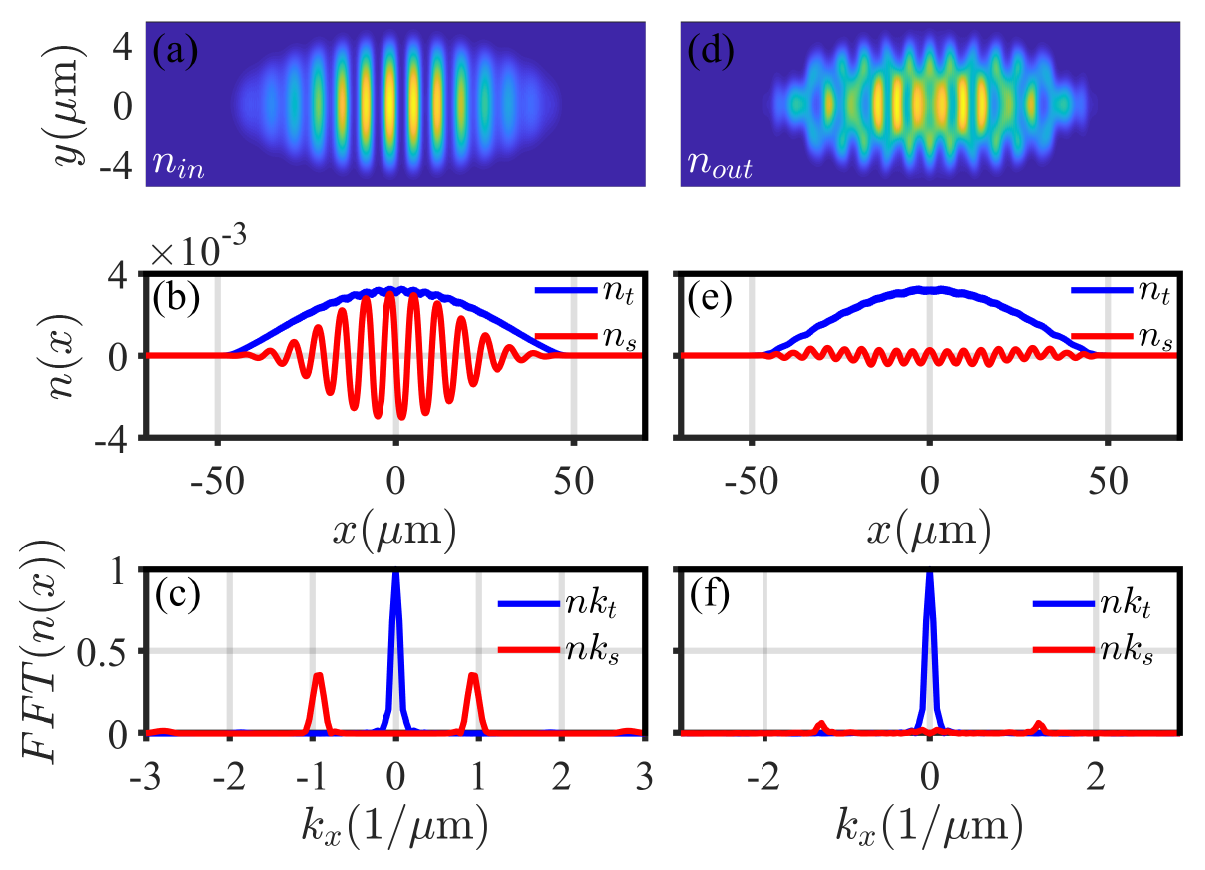}
		\caption{Comparison of in-phase (a-c) and out-of-phase (d-f) modulation results when driving frequency $\omega_i=100$ Hz in the stripe phase. Panel (a) and Panel (d) show 2D density of spin up component. Other parameters are the same as Fig. \ref{200}.}
		\label{fig3}
	\end{figure}

	Fig. \ref{200}(a) shows the Faraday patterns of each component in the stripe phase with $\Omega=0.1 E_{r}$. The scattering lengths of $^{87}$Rb atom are $a_{11}=a_{22}=100.86a_0$, $a_{12}=a_{21}=100.4a_0$, where $a_0=0.0529$ nm is the Bohr radius. In order to analyze the one-dimensional interference characteristic of the patterns, Fig. \ref{200}(b) projects the two-dimensional density onto x-direction. The red curve is $n_{s}$=$n_1-n_2$, while the blue curve is $n_{t}$=$n_1+n_2$. We can clearly observe regular oscillations in $n_{s}$ and $n_{t}$, indicating the formation of spin and density Faraday waves, respectively. Due to the coupling between $\Psi_{1}$ and $\Psi_{2}$ in the GP equations, spin and density excitation are induced simultaneously, which would be separately appear in single-component and typical two-component system. Density excitation is relatively weak as shown by the black dashed box in panel (b). 
	The presence of spin-orbit coupling results in a momentum transfer between two spin states ($\sim \hbar k_0 \sigma_{z}$), and gives rise to the spin-dependent velocity \cite{RN35A1,RN35A2,RN35A3}, $v \sim k_x-k_0 \sigma_{z}$, which is a key reason for the existence of the stripe in $n_t$ \cite{RN35A4}. Note that the blue curve in Fig. \ref{200}(b) contains two oscillations, one is clear total density oscillation in the black dash, and the other is the global small stripe. 
    Unlike Rabi-coupled two-component systems, it's possible to excite the SFW through in-phase modulation without introducing noise perturbation to the initial state \cite{RN35}. Performing a one-dimensional Fourier transform of $n_s$ and $n_t$, $FFT(n(x))$, we can get $nk_s$ and $nk_t$. The wave vectors of the red peaks are $k_{\text{spin}}\approx\pm1.3019$ $\mu$m$^{-1}$, and these of the blue peaks are $k_{\text{total}}\approx \pm2.5603$ $\mu$m$^{-1}$. Fig. \ref{200} shows the results at the modulation frequency $\omega_i=200$ Hz, which is twice the stronger trap frequency along y-direction. Next, we consider the modulation frequency $\omega_i=100$ Hz, resonating with this trap frequency. We compare the results of in-phase modulation with those of out-of-phase modulation, keeping all the other parameters consistent with Fig. \ref{200}, as shown in Fig. \ref{fig3}.

    	\begin{figure}[htbp]
		\centering
		\includegraphics[width=1\linewidth]{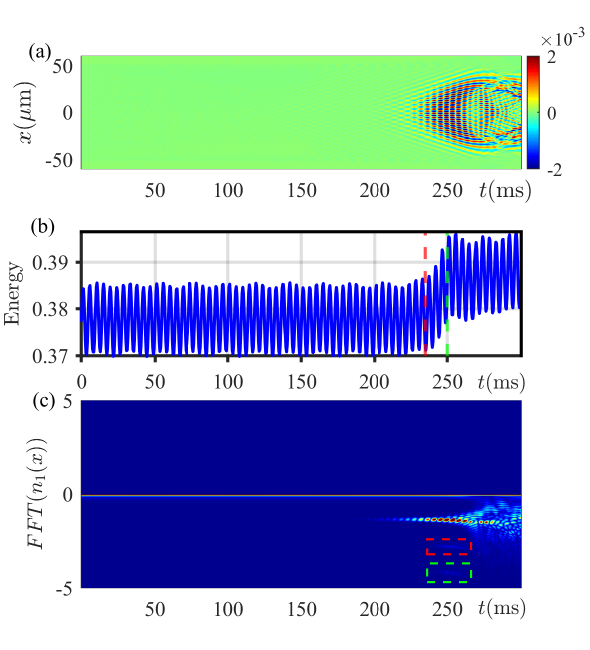}
		\caption{The temporal evolution of the system from 0 to 300 ms, resulted from in-phase modulation under the stripe phase. 
			(a) The evolution of spin density $n_s$.
			(b) The evolution of total energy before and after the formation of SFW. Red dotted line indicates 235 ms and the green one is 250 ms.
			(c) The Fourier transform of $n_1$.
			Parameters are $\Omega=0.1E_r$, $\omega_i=220$ Hz, $\Delta a=8a_0$.}
		\label{fig4}
	\end{figure}

\begin{figure}[htbp]
 	\centering
 	\includegraphics[width=1\linewidth]{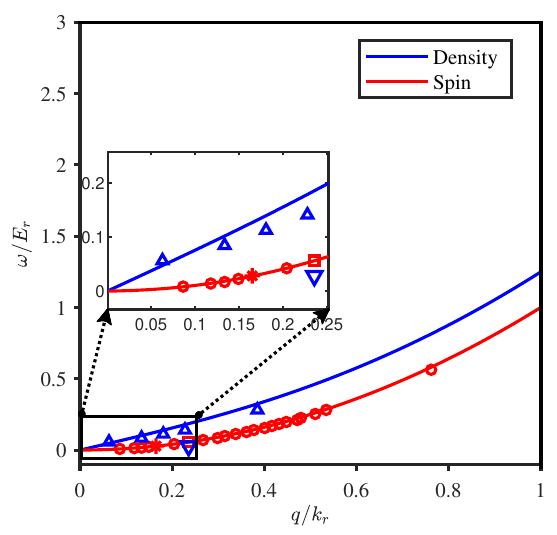}
 	\caption{Comparation of the dispersion relation between SOC and two-component BEC. The blue line and red line correspond to density and spin mode of binary BEC. The red points resulte from in-phase modulation, while blue triangles result from out-of-phase modulation. The frequency of the red square and star are $\omega_i=200$ Hz and $\omega_i=100$ Hz. The down triange is $\omega_i=100$ Hz.}
 	\label{fig5}
 \end{figure}

	The panel (a-c) of Fig. \ref{fig3} displays the results of our in-phase modulation, while panel (d-f) are the results of out-of-phase modulation based on the Eq. (\ref{out}). Firstly, we conduct investigation using out-of-phase modulation, in which the symmetry of interaction is broken. We observed the excitation pattern as a resonant wave (Fig. \ref{fig3}(d)), which corresponds to an excitation wave vector of $k_{\text{spin}}\approx\pm1.3019$ $\mu$m$^{-1}$. This result equals to the one of in-phase modulation frequency $\omega_i=200$ Hz, which is different with reference \cite{RN36,RN37}. However, when applying the in-phase modulation, we observed that the spatial pattern formed by the parameter excitation is FW (Fig. \ref{fig3}(a)) rather than resonant waves. This is a clear difference between SOC system and previous system. The points above are also confirmed quantitatively in Fig. \ref{fig5}. The patterns generated by parameter excitation also depend on the modulation method.

	Next, we analyze the formation process of SFW, as shown in Fig. \ref{fig4}. Since the initial state is the eigenstate of the Hamiltonian of the system, the collapse and spread out of condensate don’t occur during evolution \cite{RN37A1}. Fig. \ref{fig4}(a) reflects the spin dynamics by the temporal evolution of $n_s$. This longitudinal spin polarization $\langle\sigma_{z}\rangle$ is supposed to be zero before the formation of SFW, and the system remains stable during $0<t<210$ ms, suggesting that the parametric excitation strength is inadequate to produce significant phenomena. At about $t=210$ ms, a weak Spin Faraday pattern begins to appear, and gradually intensifies over tens of milliseconds. During this period, spin density n is spatially inhomogeneous and reverses with time due to the modulation \cite{RN37A2}. Finally, it evolves into granular state around $t=250$ ms \cite{RN38}, signifying the entry into the nonlinear regime. Focusing on the transverse dynamical property is reasonable because SOC exists only in x-direction. This process is also reflected clearly in Fig. \ref{fig4}(b), total energy evolution of the system over time. Before the appearance of Faraday pattern, the total energy oscillates with the modulation but remains relatively stable overall. As the Faraday pattern intensifies around $t=235$ ms, there is an increase in energy, which continues until around $t=250$ ms when the Faraday pattern disappears. In fact, after maintaining stability for a certain period of time, the total energy continues to increase, probably due to the formation of other higher order harmonics. The appearance and duration of Faraday pattern vary with modulation frequency $\omega_i$ and amplitude $\Delta a$. Fig. \ref{fig4}(c) illustrates the emergence of different modes, in which we analyze just spin up component. As there is no surface excitation, the wave vector localizes at around $-0.1$, which is result from the stripe phase. At about $t=210$ ms, a peak appears at $k_F\approx-1.389$ $\mu$m$^{-1}$, meaning the formation of the Faraday pattern. Clearer colors indicate stronger modes. At about $t=240$ ms, two additional peaks emerge at $k_R\approx-2.777$ $\mu$m$^{-1}$ (red box) and $k_H\approx-4.166$ $\mu$m$^{-1}$ (green box), corresponding to the resonant wave and a higher-order harmonic wave as discussed in \cite{RN39}. These results correspond to a certain extent to the relationship $\omega_m=2\omega_{0}/n$, where $\omega_m$, $\omega_{0}$, $n$ represent modulation frequency, intrinsic frequency and positive integer, respectively. $n=1$ correspond to the FW, and $n=2$ correspond to the resonant wave, and $n=3,...,$ correspond to higher-order harmonics.

    \begin{figure}[htbp]
 	\centering
 	\includegraphics[width=1\linewidth]{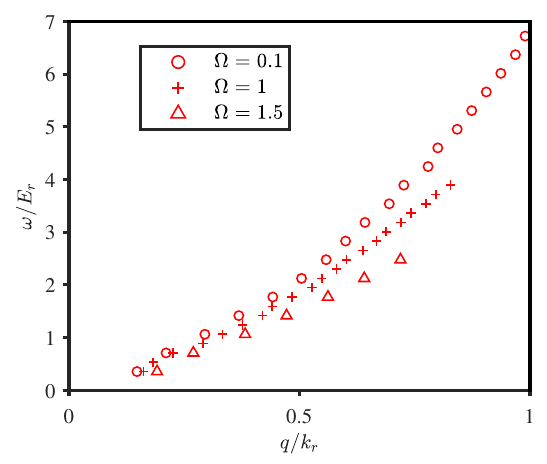}
 	\caption{Numerical simulation of the lowest branch of dispersion relation in the stripe phase with three different Rabi frequency, $\Omega=0.1E_r$, $\Omega=1E_r$, $\Omega=1.5E_r$. The maximum range of modulation frequency is $0-2000$ Hz.}
 	\label{fig6}
 \end{figure}
 
\section{ Dispersion Relation}
	\label{4}
%	\subsection{Derivation of the BdG Equation}
	
	The spinor wave function of the system is assumed as \\
	$\Psi_i(x,t)=\psi_i(x)+\delta\psi_i(x,t)$, where $\psi_i(x)$ and $\delta\psi_i(x,t)$ represent the wave function of ground state and perturbation, respectively. By substituting $\Psi_i$ directly into Eq. (\ref{gp1}) and Eq. (\ref{gp2}), and keeping only the perturbation terms, we can derive the dynamical equations that govern the excitation of the system:
	\begin{flalign}
		\begin{split}
			&i\partial_{t}\delta\psi_{1}=(\frac{\nabla^{2}-2ik_{0}\nabla-k_{0}^{2}}{2})\delta\psi_{1}+\frac{1}{2}\Omega\delta\psi_{2}+g_{11}\psi_{1}^{2}\delta\psi_{1}^{\ast} \\[2mm]
			&+2g_{11}|\psi_{1}|^{2}\delta\psi_{1}+g_{12}\psi_{1}\psi_{2}\delta\psi_{2}^{\ast}+g_{12}\psi_{1}\psi_{2}^{\ast}\delta\psi_{2}+g_{12}|\psi_{2}|^{2}\delta\psi_{1},
		\end{split}& \nonumber
	\end{flalign}

\begin{flalign}
		\begin{split}
	&i\partial_{t}\delta\psi_{2}=(\frac{\nabla^{2}+2ik_{0}\nabla-k_{0}^{2}}{2})\delta\psi_{2}+\frac{1}{2}\Omega\delta\psi_{1}+g_{22}\psi_{2}^{2}\delta\psi_{2}^{\ast} \\[2mm]
	&+2g_{22}|\psi_{2}|^{2}\delta\psi_{2}+g_{21}\psi_{2}\psi_{1}\delta\psi_{1}^{\ast}+g_{21}\psi_{2}\psi_{1}^{\ast}\delta\psi_{1}+g_{21}|\psi_{1}|^{2}\delta\psi_{2}.
		\end{split}&
	\label{ex}
	\end{flalign}
	Firstly, we consider the cases that the ground state localizes in the plane wave phase and the zero momentum phase. We assume the ground state wave function as follows:
	\begin{flalign}
&\		{\left( \begin{array}{cc}
				\psi_{1}(x,t) \\
				\psi_{2}(x,t)
			\end{array}
			\right )}
		=e^{ik_{1}x-i\mu t}\sqrt{n}
		{\left( \begin{array}{cc}
				\chi^{(0)}_{1} \\
				\chi^{(0)}_{2}
			\end{array}
			\right )},& \label{gs}
	\end{flalign}
	where $\chi^{(0)}_{1}$, $\chi^{(0)}_{2}$, $n=N/V$, $k_1$, $\mu$ denote the coefficients of wave function, average density, quasimomentum, chemical potential of the ground state, respectively. $\mu$ can be determined by 
	solving the ground state Schrödinger equation $i\partial_t\psi=(H_{\text{sp}}+H_{\text{int}})\psi$ \cite{RN40}.
	And we assume the perturbation wave function takes the form of the Bogoliubov general solution:
	\begin{flalign}
&\		\delta\psi=e^{ik_{1}x-i\mu t}\left [{\left( \begin{array}{c}
				u_{1} \\
				u_{2}
			\end{array}
			\right )} e^{iqx-iwt}+{\left( \begin{array}{ccc}
				v_{1}^{\ast} \\
				v_{2}^{\ast}
			\end{array}
			\right )}e^{-iqx+i\omega t}\right ],& \label{perturbation}
	\end{flalign}
where $u, v$ are the two Bogoliubov amplitudes, q is the perturbation quasimomentum, and $ \omega$ is the perturbation frequency. In the cases mentioned above, $u$ and $v$ are spatially independent. Substituting Eq. (\ref{gs}) and Eq. (\ref{perturbation}) into the Eq. (\ref{ex}), we obtain the BdG equation with respect to $g_{ij}$ \cite{RN41},

\begin{flalign}
&\	M{\left( \begin{array}{c}
		u_{1}\\
		u_{2}\\
		v_{1}\\
		v_{2}
		\end{array} \right )} 
		=\omega{\left( \begin{array}{c}
		u_{1}\\
		u_{2}\\
		v_{1}\\
		v_{2}
		\end{array} \right )},&  \label{eigen equation}
\end{flalign}
where M is a 4 $\times$ 4 matrix with
\begin{flalign}
&\	M={\left( \begin{array}{cc}
		H_1 & H_2 \\[4mm]
		-H_2^{\ast} & H_3
		\end{array}
		\right)},&
\end{flalign}

\begin{figure*}
%	\begin{subequations}
	\begin{normalsize}
         \begin{flalign}
		H_1=\left(\begin{array}{cc}
			\dfrac{(k+q-k_0)^2}{2}-\mu+2g_{11}|\chi_{+}^{(0)}|^2+g_{12}|\chi_{-}^{(0)}|^2 & \dfrac{\Omega}{2}+g_{12}\chi_{+}^{(0)}\chi_{-}^{(0)}		\\[4mm]
			\dfrac{\Omega}{2}+g_{21}\chi_{+}^{(0)}\chi_{-}^{(0)} & \dfrac{(k+q+k_0)^2}{2}-\mu+2g_{22}|\chi_{-}^{(0)}|^2+g_{21}|\chi_{+}^{(0)}|^2
		\end{array}
		\right),
         \end{flalign}	
     	\end{normalsize}
%\end{subequations}
\end{figure*}

\begin{figure*}
	\begin{normalsize}
%	\begin{subequations}
         \begin{flalign}
         	\setlength{\arraycolsep}{1.2pt}
			H_3=\left( \begin{array}{cc}
					-\dfrac{(k_1-q-k_0)^2}{2}+\mu-2g_{11}|\chi_{1}^{(0)}|^2-g_{12}|\chi_{2}^{(0)}|^2 & -\dfrac{\Omega}{2}-g_{12}{\chi_{1}^{(0)}}^\ast \chi_{2}^{(0)}\\[4mm]
					-\dfrac{\Omega}{2}-g_{21}\chi_{1}^{(0)}{\chi_{2}^{(0)}}^\ast &
					-\dfrac{(k_1-q+k_0)^2}{2}+\mu-2g_{22}|\chi_{2}^{(0)}|^2-g_{21}|\chi_{1}^{(0)}|^2
				\end{array}
			\right).
        \end{flalign}
    \end{normalsize}
%	\end{subequations}
{\noindent} \rule[-10pt]{18cm}{0.05em}
\end{figure*}

\begin{flalign}
	&\	H_2=\left( \begin{array}{cc}
		g_{11}{\chi_{1}^{(0)}}^{2} \qquad g_{12}\chi_{1}^{(0)}\chi_{2}^{(0)}\\[2mm]
		g_{21}\chi_{1}^{(0)}\chi_{2}^{(0)} \qquad g_{22}{\chi_{2}^{(0)}}^2
	\end{array}
	\right),&
\end{flalign}

\begin{figure}[htbp]
		\centering
        \includegraphics[width=1\linewidth]{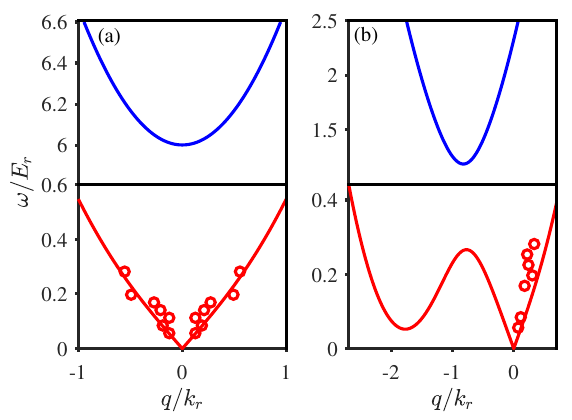}
		\caption{The two lowest branches of dispersion relation in the zero momentum phase (a) $\Omega=6 E_r$ and plane wave phase (b) $\Omega=1E_r$. Blue line and red line obtained by diagonalizing the coefficient matrix of Eq. (\ref{eigen equation}). The red points are numerical results excited by in-phase modulation with driving frequency $\omega_i$ from 200 Hz to 1000 Hz.}
		\label{fig7}
	\end{figure}	

\subsection{Stripe phase} 	
 We consider the system in the stripe phase with experimental atom parameters, where the range of stripe phase is small, $0<\Omega_{\text{stripe}}<0.19E_r$ [16]. The Raman coupling strength between the two fine levels is weak at $\Omega=0.1E_r$, and the lowest two branches of dispersion relation are close to that of traditional two-component BEC \cite{RN14,RN42}. Our numerical results strongly support this in Fig. \ref{fig5}.
 For the stripe phase, patterns can be generated clearly at smaller modulation amplitudes compared to the plane wave and zero momentum phases. In this study, we use a series of discrete values of the modulation frequency ranging from 0 to 1000 Hz and excite the lowest two energy bands of the stripe phase by applying out-of-phase and in-phase modulation. 
 Using the typical modulation frequencies, 100 Hz and 200 Hz, as examples. The excitation mode as a resonant wave is excited by out-of-phase modulation at 100 Hz. This mode exhibit the same wave vector as the FW excited by the in-phase modulation at 200 Hz, as indicated by the down triangle and red square in Fig. \ref{fig5}. In contrast, the mode excited by in-phase modulation at 100 Hz is a Faraday wave, as illustrated by the red star in Fig. \ref{fig5}. This result illustrates a difference between the SOC system and the non-SOC system.

 In our work, in order to study the variation of the dispersion relation with the Raman coupling strength $\Omega$ in the stripe phase, we set $a_{11}=a_{22}=100a_0$ and $a_{12}=a_{21}=60a_0$, which results in an increased difference in interaction strength between $g_{ii}$ and $g_{ij}$. This operation expands the range of the stripe phase to $0<\Omega_{\text{stripe}}<2E_r$. Keeping other parameters invariant, we studied the dispersion relation of the lowest branch through in-phase modulation at $\Omega=0.1E_r$, $\Omega=1E_r$ and $\Omega=1.5E_r$. The results displayed in Fig. \ref{fig6}, showed a significant reduction of the dispersion relation with the increase of $\Omega$.
 
 \subsection{Plane wave phase and zero momentum phase}

	In the case of zero momentum phase with $\Omega=6E_r$, we can find that the ground state condenses to zero, i.e., $k_1=0$ in Eq. (\ref{gs}). Minimizing the ground state energy gives $\left(\chi^{(0)}_{1}, \chi^{(0)}_{2}\right)=\left(\sqrt{2}/2, -\sqrt{2}/2\right)$ \cite{RN20}. Substituting $\mu_3$ and $\chi^{(0)}$ into Eq. (\ref{eigen equation}) and diagonalizing the coefficient matrix result in the dispersion relation as shown in Fig. \ref{fig7}(a).

	By using in-phase modulation, we simulate the dispersion relation of the FW with $0<\omega_i<1000$ Hz, $\Delta a=20a_0$. The results are these red points drawn in Fig. \ref{fig7}(a), which are in good agreement with the analytical result. We simulate only the lowest branch since our modulation energy is too low to excite the upper branch in the presence of the energy gap.
	
	In the plane wave phase at $\Omega=1E_r$, we prepare an initial state with quasimomentum $+k_1$. This state corresponds to the assumption of Eq. (\ref{gs}), and the system localizes at $k_1=k_0\sqrt{1-\Omega^2/[2(k^2_0-2G_2)]^2}$, $G_2=n(g_{ii}-g_{ij})/4$. At this point, minimizing the ground state energy gives $\left(\chi^{(0)}_{1},\, \chi^{(0)}_{2}\right)=\left(\sqrt{\dfrac{1}{2}+\dfrac{k_1}{2k_0}},\, -\sqrt{\dfrac{1}{2}-\dfrac{k_1}{2k_0}}\right)$. Using the same method as above, substituting $\mu_2,\, \chi^{(0)},\, k_1$ into Eq. (\ref{eigen equation}), we get the dispersion shown in Fig. \ref{fig7}(b).

    \section{Conclusion}
	\label{5}
	
	We mainly explore the SFW in a two-dimensional SOC BEC by modulating the scattering lengths periodically. A converse momentum shift between two components is induced in the stripe phase, which breaks the translational symmetry and U(1) symmetry. SFW in this phase can be excited with in-phase modulation and without noise. We observe the formation of SFW over a large frequency range of $0-4000$ Hz.
	In this process, we discover an intriguing phenomenon: under the stripe phase with in-phase modulation, when the modulation frequency equals to the stronger confinement frequency, the excitation remains FW rather than resonant waves \cite{RN37,RN39}, as shown in Fig. \ref{fig5}. This result is significantly different from both single-component \cite{RN5,RN11,RN37} and two-component \cite{RN43} system. 
    Furthermore, we study the evolution of the system over time at a general frequency parameter $\omega_i=220$ Hz in the stripe phase. We find that Faraday pattern remains stable for about 50 ms and are accompanied by the generation of resonant waves and higher order harmonics, after which the system enters the granular state. We conduct numerical simulation of the dispersion relation in stripe phase and the other two phases using periodic modulation instead of the commonly used Bragg spectroscopy method in experiment \cite{RN44}. The dispersion relation in stripe phase matches well with traditional two-component BEC when Rabi coupling is weak. And the simulation results in the plane wave phase and zero momentum phase are in good agreement with BdG equations. Our research results are also applicable to three-dimensional situations.
    
    In our study, we find the quadrupole modes are excited at some specific modulation frequencies in the stripe phase, which are similar to the zero momentum phase studied in \cite{RN45}. 
    Due to the supersolid-like characteristic of SOC BEC in the stripe phase, it would be an interesting question to study the dependence of the spin excitation of periodically modulated stripe phase on various experimental parameters, including the modulation amplitude and frequency.
    
\begin{acknowledgments}
	
H.L., J.W., and Y. L. are supported by the National Natural Science Foundation of China (Grant No. 11774093 and No. 12074120), the Natural Science Foundation of Shanghai (Grant No. 23ZR1418700), the Innovation Program of Shanghai Municipal Education Commission (Grant No. 202101070008E00099) and the Program of Chongqing Natural Science Foundation (Grant No. CSTB2022NSCQ-MSX0585).
\end{acknowledgments}

%%%%

\end{document}